\documentstyle[newarcrc,fleqn,psfig]{article}
\newcommand{\gtrsim}{\mathrel{\hbox{\rlap{\lower.55ex \hbox {$\sim$}}
                   \kern-.3em \raise.4ex \hbox{$>$}}}}
\newcommand{\lesssim}{\mathrel{\hbox{\rlap{\lower.55ex \hbox {$\sim$}}
                   \kern-.3em \raise.4ex \hbox{$<$}}}}

\title{Towards 4U\,1630$-$47: a black-hole soft X-ray transient odyssey}

\author{Erik Kuulkers\address{Astrophysics, University of Oxford, Nuclear and
Astrophysics Laboratory, \\Keble Road, Oxford OX1 3RH, United Kingdom}}

\begin{document}
% typeset front matter
\maketitle

\begin{abstract}
4U\,1630$-$47 is a black-hole X-ray transient with one of the shortest 
recurrence times. Despite its regular outburst behaviour little is known about 
this source. Only recently has attention to this system increased.
I discuss there the basic known (X-ray) properties of 4U\,1630$-$47 and 
report on X-ray and radio observations obtained during its recent outburst, 
starting in 1998 February.
These observations strengthen some of the similarities seen between
4U\,1630$-$47 and the Galactic superluminal sources GRO\,J1655$-$40 and
GRS\,1915+105, and provide the first detection of 4U\,1630$-$47 in the radio.
Using an updated outburst ephemeris I predict the next outburst to occur
about a week before Christmas 1999.
\end{abstract}

\section{Introduction}

4U\,1630$-$47 is a member of the transient class of low-mass X-ray binaries.
The nature of the compact star in the system is, however, unknown.
Its X-ray spectral (Parmar, Stella \&\ White 1986; Barret, McClintock 
\&\ Grindlay 1996) and X-ray timing (Kuulkers, van der Klis \&\ Parmar 
1997a) properties during outburst suggest it is a black-hole.
Until 1998 no optical/infra-red or radio counterpart has been reported.

Soon after its discovery in 1972, 4U\,1630$-$47 was shown to exhibit outbursts 
every $\sim$600 days (Jones et al.\ 1976; Priedhorsky 1986), i.e.\ one of the 
shortest recurrence times among the black-hole soft X-ray transients. 
It became clear that 4U\,1630$-$47 continued to undergo regular outbursts at 
approximately the 600 day 
period (Parmar, Angelini \&\ White 1995; Parmar et al.\ 1997).
However, by including the outbursts as seen with the {\it Ginga} ASM and the 
{\it RXTE} ASM it was found that the outburst recurrence interval 
changed from $\sim$600 days to $\sim$690 days, somewhere between 1984 and 1987
(Kuulkers et al.\ 1997b). 
Such a large change in recurrence time suggests that the 600/690~day period 
is not associated with the binary period of the system.
It was predicted that if the recurrence interval of $\sim$690~days continued, 
the next outburst of 4U\,1630$-$47 would occur near 1998 Jan 31
(JD\,245\,0845).

The outburst behaviour of 4U\,1630$-$47 is more complex than was previously 
realized. In the standard X-ray band ($\sim$2--10\,keV), the source shows 
outbursts with durations on the order of a few months (e.g.\ Fig.~1) and 
sometimes intervals 
of long-term X-ray activity of up to $\sim$2.4~years (Kuulkers et al.\ 1997b). 
X-ray activity in the hard X-ray band ($\sim$20--100\,keV) associated with
these outbursts has also been seen by {\it CGRO} BATSE (Bloser et al.\ 1996,
1998), but lasted for only $\sim$weeks.
The detection of \mbox{X-ray} absorption dips during outburst with the
{\it RXTE} PCA (Tomsick et al.\ 1998; Kuulkers et al.\ 1998a) suggests that we
see the system at an inclination angle of $\sim$60--75$^{\circ}$.

Recently, Kuulkers et al.\ (1997a,b; 1998a) pointed out 
similarities in the X-ray behaviour between 4U\,1630$-$47 and the two
Galactic superluminal X-ray transient sources GRO\,J1655$-$40 and 
GRS\,1915+105. It was, therefore, postulated that the physics involved
are similar in these systems, and that 4U\,1630$-$47 might also show 
radio activity during its outburst, possibly in the form of (relativistic) 
jets.

\section{1998 Outburst}

\subsection{X-ray observations: showing the strongest QPO ever seen}

Both the {\it RXTE} ASM and {\it CGRO} BATSE picked up renewed X-ray activity 
from 4U\,1630$-$47 near 1998 Feb 2 (JD\,245\,0847), i.e.\ very close to 
that predicted (see Sect.~1; Kuulkers et al.\ 1998b; Hjellming et al.\ 1998). 
The source did not rise to maximum immediately, but exhibited a 
standstill in the 2--12\,keV band for about a week. During this time 
{\it CGRO} BATSE showed a considerable contribution from 4U\,1630$-$47,
indicating that during this initial part of the outburst the source 
spectrum was hard.

The 1998 outburst as seen with the {\it RXTE} ASM is shown in 
Fig.~1b. The shape of the outburst differs considerably from the previous 
outburst in 1996, which was covered as well by the {\it RXTE} ASM (Fig.~1a). 

\begin{figure}
\centerline{
\psfig{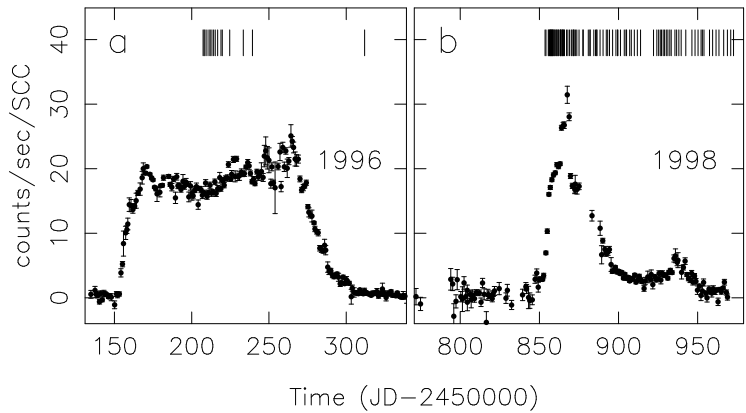}}
\caption{{\it RXTE} ASM outburst light curve of 4U\,1630$-$47 in 1996
({\bf a}) and 1998 ({\bf b}) on the same scale. Plotted are the daily averages 
as obtained from
the MIT {\it RXTE} ASM web pages. $\sim$75\,cts\,s$^{-1}$\,SCC$^{-1}$
corresponds to 1~Crab. Note the clear differences in outburst shape.
Indicated at the top are the times of the pointings with the 
{\it RXTE} PCA/HEXTE instruments.}
\end{figure}

As soon as 4U\,1630$-$47 started to rise rapidly as seen with the 
{\it RXTE} ASM, TOO programs were triggered with the 
{\it RXTE} PCA/HEXTE. Since Feb 9 (JD\,245\,0854) almost daily pointings 
were obtained (see top of Fig.~1b), in order to investigate the spectral and 
timing behaviour of the source as it evolved through possible different black-hole transient 
states. Clearly, this outburst has been well covered, compared to e.g.\ the 
1996 outburst (Fig.~1a).
 Several other X-ray satellites also performed TOO observations, 
including {\it ROSAT} HRI/PSPC between Feb 17 and 28
(JD\,245\,0862--873), BeppoSAX (Oosterbroek et al.\ 1998) on Feb 20, 24, 
Mar 7/8, 19/20 and 26/27 (JD\,245\,0865, 871, 880/881, 892/892, 899/900) and 
ASCA on Feb 25/26 (JD\,245\,0871/872).

The first two {\it RXTE} PCA observations on Feb 9 
showed two strong quasi-periodic oscillations 
(QPO) peaks and strong band-limited noise in the power spectrum of the 
2--14\,keV X-ray intensity (Dieters et al.\ 1998).
During the first observation the lowest-frequency QPO peak was 
asymmetric and had a centroid frequency of $\sim$2.7\,Hz, 
FWHM of $\sim$0.2\,Hz and a fractional 
rms amplitude of $\sim$16\%\ ($\sim$23\%\ including the asymmetric wing), 
whereas for the highest-frequency QPO the values were 
$\sim$5.6\,Hz, 1.2\,Hz and $\sim$5.5\%, respectively.
The higher-frequency QPO is most likely the harmonic of the 
lower-frequency QPO.
The QPO near 3\,Hz is the strongest ever seen (see e.g.\ van der Klis
1995). 

In subsequent {\it RXTE} PCA observations (Dieters et al., in preparation), 
during which the X-ray intensity 
increased further, the QPO centroid frequency increased as well; it 
became also broader, and its strength decreased. When the {\it RXTE} ASM
flattened off (near Feb 14; JD\,245\,0859), the 3--10\,Hz QPO had 
vanished. The fast increase near Feb 18 (JD\,245\,0863) marked the beginning
of the so-called high state (HS), generally seen in black-hole binaries.

Dieters et al.\ (1998) noted that the power spectral shape showed 
similarities to those seen in power spectra in the so-called very-high state
(VHS) of black-hole binaries (see e.g.\ van der Klis 1995). However, in the 
VHS the $\sim$1--10\,keV X-ray intensity is generally high (even higher than 
in the HS).
Instead, the 3--10\,Hz QPO during the low soft and high hard X-ray intensity
state of 4U\,1630$-$47, resembles the QPO seen in GRS\,1915+105 in a similar 
frequency range and also during one of its similar low soft and high hard X-ray 
intensity state (so-called ``low-hard" state; 
Morgan, Remillard \&\ Greiner 1997).

\subsection{Radio observations: tune in to 4U\,1630$-$47}

\begin{figure}
\centerline{
\psfig{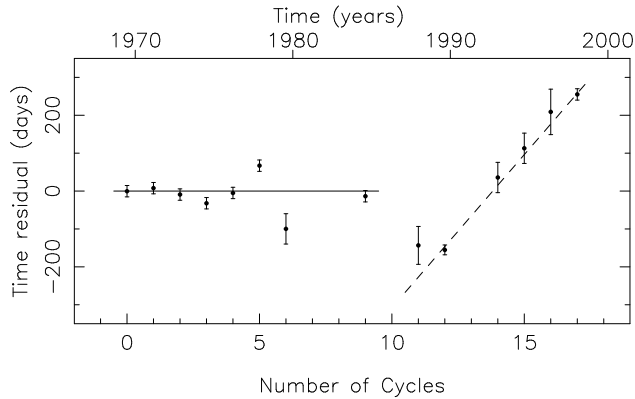}}
\caption{The time residual in days of all outburst times after a linear fit
to the first eight outburst times to cycle number. The error bars reflect
the uncertainties in deriving the outburst start times (see text; 
Parmar et al.\ 1997; Kuulkers et al.\ 1997b). The dashed line shows the 
linear fit to the last six residual outburst times to cycle number.
As shown by Kuulkers et al.\ (1997b), the recurrence interval changed its 
period and phase between the 1984 (cycle 9) and 1987 outbursts (cycle 11).}
\end{figure}

Radio measurements when the 1996 outburst was well underway failed to detect 
emission from 4U\,1630$-$47, with 3$\sigma$ upper limits of 5\,mJy and 2\,mJy
at 1.4\,GHz and 2.4\,GHz, respectively (Kuulkers et al.\ 1997b).
The first radio observations of the 1998 outburst on Feb 9
also did not show any emission ($<$0.7\,mJy at 8.6\,GHz; 
Kuulkers et al.\ 1998b). However, a few days later (Feb 12; JD\,245\,0857) a 
radio source was discovered with the VLA within the X-ray position error 
region (Hjellming \&\ Kuulkers 1998). Subsequently, the source was confirmed by 
ATCA (Buxton et al.\ 1998).
The radio observations allowed Hjellming et al.\ (1998) to identify an 
accurate position of 4U\,1630$-$47 (equinox 2000.0): 
16$^h$34$^m$1.60$^s$$\pm$0.05$^s$, $-$47$^{\circ}$23'33"$\pm$2" (VLA)
or 16$^h$34$^m$1.60$^s$$\pm$0.02$^s$, $-$47$^{\circ}$23'34.8"$\pm$0.3"
(ATCA).

The radio outburst most probably started between Jan 30 and Feb 5 
(JD\,245\,0844--850), i.e.\ during which the source was in a hard state
(Hjellming et al.\ 1998).
The radio outbursts seen in GRO\,J1655$-$40 and GRS\,1915+105, and possibly 
in other soft X-ray transients, have been
seen to originate during similar hard X-ray states 
(see Hjellming et al.\ 1998, and references therein).
The peak of the radio outburst occurred near Feb 18 (JD\,2450862) with 
a flux density of $\sim$2.6\,mJy at 4.8\,GHz
(Buxton et al.\ 1998; Hjellming et al.\ 1998). During these peak measurements
the ATCA detected very strong linear polarisation, i.e.\ $\sim$28\%\ at
4.8\,GHz. This is much larger than is generally found in radio sources 
($\lesssim$10\%\/); large linear polarization ($\gtrsim$16\%\/) has up to now
only been seen from jet sources like SS433, GRO\,J1655$-$40 and GRS\,1915+105
(Hjellming et al.\ 1998). As suggested by Hjellming et al.\ the radio
(and correlated X-ray) properties might imply that 4U\,1630$-$47 
ejected radio jets during its 1998 outburst.

\section{Outburst ephemeris}

The 1998 outburst of 4U\,1630$-$47 enables me to update the outburst 
ephemeris. For the time of occurrence of
the 1998 outburst I take the time of peak intensity from the {\it RXTE} ASM 
data, i.e.\ near Feb 22 (JD\,245\,0867). The peak is well determined;
I use, however, a typical uncertainty of 15~days, 
similar to e.g.\ Parmar et al.\ (1997).

For the fit to the outburst recurrence times I follow Kuulkers et al.\ (1997b),
where they allowed the recurrence period and phase to change between the
1984 and 1987 outbursts. The resulting period before the period/phase change 
is still 601$\pm$2~days; after the change it is 
now 682$\pm$4~days. The deviation of the outburst times from the expected 
times of
a $\sim$601~day period is shown in Fig.~2, together with the results of the fits.
Assuming that the next outburst occurs $\sim$682~days after the start of the 
1998 outburst, I predict it will occur about a week before Christmas 1999,
i.e.\ near 1999 December 16 (JD\,245\,1529). 

\section*{Acknowledgements}

I would like to give loads of credit to T.~Augusteijn, T.~Belloni, M.~Buxton,
S.~Dieters, T.~Dotani, B.~Hjellming, M.~McCollough, T.~Oosterbroek, A.~Parmar
and R.~Sood, without whom this report and/or much of my work on 4U\,1630$-$47 
could not have been possible.

\end{document}